\documentclass[
	,smallextended
]{svjour3}

\usepackage{array}
\usepackage{booktabs}
\usepackage{float}
\usepackage{mathtools}
\usepackage{xcolor}
\usepackage[
	,a4paper
	,truedimen
	,nomarginpar
	,headsep=.020\paperheight
	,foot=.042\paperheight
	,vscale=.84
	,hscale=.80
	,vmarginratio=7:5
	,hcentering
	,mag=1414
]{geometry}
\usepackage[T1]{fontenc}
\usepackage{textcomp}
\usepackage[utf8]{inputenc}
\usepackage{lmodern}
\usepackage{libertineRoman}
\usepackage[
	,slantedGreek
	,libertine
	,libaltvw
	,liby
]{newtxmath}
\usepackage{physics}

\usepackage{mleftright}\mleftright
\usepackage{upgreek}
\usepackage{bm}
\usepackage[
	,colorlinks
	,allcolors=blue!90!cyan!80!black!90!white
	,setpagesize=false
	,bookmarksnumbered=true
	,bookmarksopen=true
	,pdfborder={0 0 0}
]{hyperref}
\newcommand{\br}{{\bm r}}
\newcommand{\bp}{{\bm p}}
\newcommand{\cI}{\mathcal{I}}
\newcommand{\veps}{\varepsilon}
\newcommand{\ui}{\mathrm{i}}
\newcommand{\ue}{\mathrm{e}}

\newcommand{\xa}{\alpha}

\newcommand{\F}{\mathrm{F}}

\newcommand{\B}{\mathrm{B}}
\newcommand{\BB}{\mathrm{BB}}
\newcommand{\BF}{\mathrm{BF}}
\newcommand{\FB}{\mathrm{FB}}
\newcommand{\BEC}{\mathrm{BEC}}

\newcommand{\TF}{T_\mathrm{F}}

\newcommand{\ideal}{\mathrm{ideal}}

\newcommand{\kB}{k_{\mathrm B}}

\newcommand{\zomega}{\omega}
\newcommand{\mix}{\mathrm{in}}
\newcommand{\slow}{\mathrm{slow}}
\newcommand{\fast}{\mathrm{fast}}
\begin{document}
\title{
Dipole Mode of Trapped Bose--Fermi Mixture Gas
}
\author{%
Yoji~Asano, Shohei~Watabe, and Tetsuro~Nikuni
}
\institute{%
Y.~Asano, S.~Watabe, and T.~Nikuni \at
Department of Physics,
Tokyo University of Science,
1-3 Kagurazaka,
Shinjuku-ku,
Tokyo,
162-8601,
Japan
}
\date{\today}
\maketitle

\begin{abstract}
We investigate dipole modes in a trapped Bose--Fermi mixture gas in the normal phase, composed of single-species bosons and single-species fermions with $s$-wave scattering.
In the extremely low temperature regime, Bose--Einstein statistics and Fermi--Dirac statistics may give rise to an interesting temperature dependence of collective modes.
Applying the moment method to the linearized Boltzmann equation, we study the transition of the dipole modes between the hydrodynamic regime and the collisionless regime.
\end{abstract}

\section{Introduction}
Ultracold atomic gases have opened up opportunities for studying new quantum systems, which have been not addressed before.
In contrast to the liquid helium $^3\mathrm{He}$ and/or $^4\mathrm{He}$, the atomic gases provide a wide variety of mixtures, with changing quantum statistics, a mass of particles, a number of particles, a strength of interaction, and geometry of
confinement~\cite{onofrio_2016_PU_59_physics}.
In particular, Bose--Fermi mixture is one of the most interesting mixture gases, because this system involves different quantum statistics.

In mixture gases of Bose--Einstein condensates (BECs) and degenerate Fermi gases, collective modes such as quadrupole and breathing modes have been experimentally investigated~\cite{fukuhara_2009_APB_96_quadrupole,huang_2019_PRA_99_breathing}.
There also have been theoretical investigations on, for example, monopole and multipole modes using the variational-sum-rule
approach~\cite{banerjee_2007_PRA_76_collective},
and using the scaling ansatz
formalism~\cite{liu_2003_PRA_67_collisionless}, 
low-lying modes in spinor Bose--Fermi
mixtures~\cite{pixley_2015_PRL_114_damping}, 
density and single-particle excitation using the random-phase
approximation~\cite{capuzzi_2001_PRA_64_zero},
monopole and multipole modes for trapped phase-segregated
mixtures~\cite{schaeybroeck_2009_PRA_79_trapped}, 
and dipole modes using the variational-sum-rule
approach~\cite{banerjee_2009_JPBAMOP_42_dipole}.

We consider a trapped Bose--Fermi mixture gas composed of single-species bosons and single-species fermions.
All the interactions are described by $s$-wave scattering, where collisions between fermions are prohibited by Pauli blocking.
Pauli blocking may suppress the interspecies (Bose--Fermi) collisions, while collisions between bosons may be enhanced by Bose enhancement.
These quantum statistical properties may provide an interesting feature of the collective modes.
In fact, in the case of a spatially uniform system, we argued that there exists a long-lived sound mode between the collisionless regime and the hydrodynamic regime~\cite{asano_2019_JLTP_196_collective}.
In this study, we explore exotic behaviors in dipole modes which may be observed in harmonically trapped gases.
We focus on the normal phase, where neither the BEC nor the Cooper pairs are present, with employing the moment method to the linearized Boltzmann equation.
This method is useful for analyzing dynamics of dilute gases, because it can describe a transition between the hydrodynamic regime and the collisionless 
regime~\cite{guery-odelin_1999_PRA_60_collective,nikuni_2002_PRA_65_finite,ghosh_2000_PRA_63_collective,watabe_2010_JLTP_158_zero,narushima_2018_JPBAMOP_51_density}.

\section{Moment method}
We start with the Boltzmann equation as
\begin{align}&
\pdv{f_\xa}{t}+\pdv{\veps_\xa}{\bp}\cdot\pdv{f_\xa}{\br}
-\pdv{U_\xa}{\br}\cdot\pdv{f_\xa}{\bp}=\cI_\xa
\;,
\label{eq:Boltzmann_equation}%
\end{align}
where $\xa=\{\B,\F\}$ represents bosons and fermions, respectively.
Here, $f_\xa=f_\xa(\br,\bp,t)$ is a distribution function, and the single-particle energy $\veps_\xa=\veps_\xa(\br,\bp,t)$ is given by 
\begin{align}&
\veps_\xa(\br,\bp,t)=\frac{\bp^2}{2m_\xa}+U_\xa
\;,
\label{eq:single-particle_energy}%
\end{align}
where $m_\xa$ is an atomic mass.
The potential energy $U_\xa=U_\xa(\br,t)$ is a sum of the mean-field term and the trapping potential term, given by 
\begin{subequations}
\label{eq:trap_potentials}%
\begin{align}&
U_\B(\br,t)=2g_\BB n_\B+g_\BF n_\F+\frac{m_\B\zomega_\B^2}{2}{\br}^2
\;,\\&
U_\F(\br,t)=g_\BF n_\B+\frac{m_\F\zomega_\F^2}{2}{\br}^2
\;,
\end{align}
\end{subequations}
where $n_\xa=\int\dd[3]pf_\xa/(2\uppi\hbar)^3$ is the number density, $\zomega_\xa$ is a harmonic trap frequency, and $g_\BB=4\uppi\hbar^2a_\BB/m_\B$ and $g_\BF=2\uppi\hbar^2 a_\BF/m_\BF$ with the reduced mass $m_\BF=m_\B m_\F/(m_\B+m_\F)$ are coupling strengths of the Bose--Bose and the Bose--Fermi interactions, respectively.
Here, $a_\BB$ and $a_\BF$ are Bose--Bose and Bose--Fermi scattering lengths, respectively.
Since Bose--Fermi mixture we consider is composed of single-species bosons and single-species fermions with $s$-wave scattering, Pauli blocking prohibits the Fermi--Fermi interactions.

The collision integrals $\cI_{\B,\F}=\cI_{\B,\F}(\br,\bp,t)$ in the present case are given by
\begin{subequations}
\begin{align}&
\cI_\B
=\cI_\BF[f_\B,f_\F]+\cI_\BB[f_{\B,\B}]
\;,\\& 
\cI_\F
=\cI_\FB[f_\F,f_\B]
\;.
\end{align}
\end{subequations}
The collision integral $\cI_{\xa\beta}$ for $\{\xa,\beta\}=\{\B,\B\}$, $\{\B,\F\}$, or $\{\F,\B\}$ is defined by
\begin{align}&
\cI_{\xa\beta}
= 
A_{\xa\beta} 
\int\!\!\!\frac{\dd[3]{p_2}}{(2\uppi\hbar)^3}
\int\!\!\!\frac{\dd[3]{p_3}}{(2\uppi\hbar)^3}
\int\!\!\!\dd[3]{p_4}
\delta_\bp(1234)
\delta_E^{\xa\beta}(1234)
F_{\xa\beta}(1234)
\;, 
\end{align}
where $A_\BF=A_\FB=2\uppi g_\BF^2/\hbar$, $A_\BB=4\uppi g_\BB^2/\hbar$, and we have used
\begin{subequations}
\begin{align}&
\delta_\bp(1234)
=  
\delta(\bp_1+\bp_2-\bp_3-\bp_4)
\;,\\& 
\delta_E^{\xa\beta}(1234)
= 
\delta\pqty\big{\veps_\xa(\bp_1)+\veps_\beta(\bp_2)-\veps_\beta(\bp_3)-\veps_\xa(\bp_4)}
\;,\\&
F_{\xa\beta}(1234)
=\bqty\big{1+\eta_\xa f_\xa(1)}\bqty\big{1+\eta_\beta f_\beta(2)}
f_\beta(3)f_\xa(4)
\notag\\&\hphantom{F_{\xa\beta}(1234){}={}} 
-
f_\xa(1)f_\beta(2)
\bqty\big{1+\eta_\beta f_\beta(3)}\bqty\big{1+\eta_\xa f_\xa(4)}
\;.
\end{align}
\end{subequations}
Here, we have taken $\eta_\B=+1$ and $\eta_\F=-1$, depending on the quantum statistics.

To analyze the collective oscillations, we consider a deviation of the distribution function $\delta f_\xa$ from the equilibrium distribution function $f^0_\xa$, given by $\delta f_\xa=f_\xa-f_\xa^0$, where $f^0_\xa=\Bqty\big{\ue^{\bqty*{p^2/(2m_\xa)+U^0_\xa-\mu_\xa}/(\kB T)}-\eta_\xa}^{-1}$ is the Bose--Einstein distribution or Fermi--Dirac distribution function.
In this function, the equilibrium effective potential $U^0_\xa$ is given
by Eq.~\eqref{eq:trap_potentials}
with the equilibrium number density $n_\xa^0$.
From the linearized form of the Boltzmann
equation~\eqref{eq:Boltzmann_equation},
one can derive the equation of motion for the average of an arbitrary dynamical quantity $\chi=\chi(\br,\bp)$ as 
\begin{align}&
\dv{\delta\ev{\chi}_\xa}{t}
-\frac1{m_\xa}\delta\ev{\bp\cdot\pdv{\chi}{\br}}_\xa
+\delta\ev{\pdv{U_\xa^0}{\br}\cdot\pdv{\chi}{\bp}}_\xa
-
\ev{
\chi
\pdv{f^0_\xa}{\veps^0_\xa}\frac\bp{m_\xa}
\cdot
\pdv{\delta U_\xa}{\br} 
}
=\ev{\chi\cI_\xa}
\;.
\label{eq:equation_of_motion_for_the_average}%
\end{align}
Here, we have defined the following averages:
\begin{subequations}
\begin{align}&
\delta\ev{\chi}_\xa
=\frac1{N_\xa}\iint\frac{\dd[3]{r}\dd[3]{p}}{(2\uppi\hbar)^3}
\chi(\br,\bp)\delta f_\xa(\br,\bp,t)
\;,\\&
\ev{\chi A_\xa}
=\frac1{N_\xa}\iint\frac{\dd[3]{r}\dd[3]{p}}{(2\uppi\hbar)^3}
\chi(\br,\bp)A_\xa(\br,\bp,t)
\;,
\end{align}
\end{subequations}
where $N_\xa$ is the total number of particles for $\xa=\{\B,\F\}$, and $A_\xa$ is an arbitrary function of $\br$, $\bp$, and $t$.
We have introduced the fluctuation of the potential energy $\delta U_\B=2g_{\BB}\delta n_\B+g_\BF\delta n_\F$, and $\delta U_\F=g_\BF\delta n_\B$.

A dipole mode is described by a displacement of the center of mass.
From Eq.~\eqref{eq:equation_of_motion_for_the_average}, 
we obtain coupled moment equations, where the displacement of the center of mass is along $z$-direction:  
\begin{subequations}
\label{eq:dipole_moment_eq}%
\begin{align}&
\dv{\delta\ev{z}_\xa}{t}-\frac1{m_\xa}\delta\ev{p_z}_\xa
=0
\;,
\label{eq:dipole_moment_eq_a}
\\&
\dv{\delta\ev{p_z}_\xa}{t}
+\pqty{
m_\xa\zomega_\xa^2
-\frac{\Delta}{N_\xa}
}\delta\ev{z}_\xa
+\frac{\Delta}{N_\xa}\delta\ev{z}_\beta
=-\frac{M_+}{\tau N_\xa}
\pqty{\frac{\delta\ev{p_z}_\xa}{m_\xa}-\frac{\delta\ev{p_z}_{\beta}}{m_{\beta}}}
\;,
\label{eq:dipole_moment_eq_b}
\end{align}
\end{subequations}
where $\{\xa,\beta\}=\{\B,\F\}$, $\{\F,\B\}$, the total reduced mass $1/{M_\pm}=1/{M_\B}\pm1/{M_\F}$ with $M_\xa=m_\xa N_\xa$.
Here, the mean-field contribution $\Delta$ depending on spatial profiles of the number densities is given by 
\begin{align}& 
\Delta
=g_\BF\int\!\!\!\dd[3]{r}
\pdv{n^0_\B}{z}\pdv{n^0_\F}{z}
\;,
\end{align}
and the relaxation time $\tau$ originated from the interspecies scattering is given by
\begin{align}&
\frac1{\tau}
=\frac{3\uppi\beta g_\BF^2}{\hbar M_+}
\int\!\!\!\dd[3]{r}
\int\!\!\!\frac{\dd[3]{p_1}}{(2\uppi\hbar)^3}
\int\!\!\!\frac{\dd[3]{p_2}}{(2\uppi\hbar)^3}
\int\!\!\!\frac{\dd[3]{p_3}}{(2\uppi\hbar)^3}
\int\!\!\!\dd[3]{p_4} 
\delta_\bp(1234)
\delta_E^\BF(1234)
\notag\\&\hphantom{\frac1{\tau}{}={}\frac{3\uppi\beta g_\BF^2}{\hbar M_+}}
\times 
\bqty\big{1+f^0_\B(1)}\bqty\big{1-f_\F^0(2)}f_\F^0(3)f^0_\B(4)
\pqty{p_{1z}-p_{4z}}^2
\;.
\end{align}
We note that these temperature-dependent quantities only have explicit dependence of the interspecies interaction ($\Delta\propto g_\BF$ and $1/\tau\propto g_\BF^2$).
In deriving Eq.~\eqref{eq:dipole_moment_eq_b}
in a closed form, we have truncated the mean-field terms and the collision terms by approximating that the velocity field is independent of position.

By considering the normal-mode solution $\delta\ev{\chi}_{\B,\F}\propto\ue^{-\ui\omega t}$, 
we obtain quartic equation for $\omega$ that determines eigenfrequencies of dipole modes.
In the hydrodynamic limit $\zomega_\xa\tau\ll1$, we obtain the solutions
\begin{align}&
\omega
=
\begin{dcases}
\Omega_\mix-\ui\Gamma_\mix
\;,\\ 
\hphantom{\omega_\mix}
-\ui\Gamma_\fast
\;,\\
\hphantom{\omega_\mix}
-\ui\Gamma_\slow 
\;, 
\end{dcases}
\label{eq:eigenvalue_in_hydrodynamic_limit_in_dipole_mode}
\end{align}
where 
\begin{subequations}
\begin{align}&
\Omega_\mix^2
=\frac{M_\B\zomega_\B^2+M_\F\zomega_\F^2}{M_\B+M_\F}
\label{eq:frequency_of_in-phase_mode}
\;,\\&
\Gamma_\mix
=\tau\frac{M_+\pqty\big{\zomega_\B^2-\zomega_\F^2}^2}{2\pqty\big{M_\B\zomega_\B^2+M_\F\zomega_\F^2}}
\;,\\&
\Gamma_\fast
=\frac1{\tau}
\;,\\&
\Gamma_\slow
=\tau\pqty{\frac{\zomega_\B^2\zomega_\F^2}{\Omega_\mix^2}-\frac{\Delta}{{M_+}}}
\;.
\end{align}
\end{subequations}
In the special case $\zomega_\B=\zomega_\F\,(\equiv\zomega_0)$, one can easily find that $\Omega_\mix=\zomega_0$ and $\Gamma_\mix=0$, which corresponds to the well-known Kohn 
mode~\cite{kohn_1961_PR_123_cyclotron}
(the undamped dipole oscillation independent of interactions, temperature, and quantum statistics).
In the collisionless limit $\zomega_\xa\tau\gg1$, we obtain solutions $\omega=\Omega_+-\ui\Gamma_+$ and $\omega=\Omega_--\ui\Gamma_-$.
The frequency $\Omega_\pm$ and the damping rate $\Gamma_\pm$ are given by
\begin{subequations}
\label{eq:eigenfrequency_and_damping_rate_in_collisionless_limit_in_dipole_mode}%
\begin{align}&
\Omega_\pm^2= 
\omega_+^2-\frac{\Delta}{2M_+}
\pm
\sqrt{%
\omega_-^2
\pqty{\omega_-^2-\frac{\Delta}{M_-}}
+\frac{\Delta^2}{4M_+^2}
}
\;, 
\label{eq:eigenfrequency_in_collisionless_limit_in_dipole_mode}
\\&
\Gamma_\pm 
=  
\frac1{4\tau}
\bqty{%
1\pm 
\frac1{\Omega_+^2-\Omega_-^2}
\pqty{\frac{2M_+}{M_-}\omega_-^2
-\frac{\Delta}{M_+}}
}
\;,
\label{eq:damping_rate_in_collisionless_limit_in_dipole_mode}
\end{align}
\end{subequations}
where $\omega_\pm^2=\pqty\big{\zomega_\B^2\pm\zomega_\F^2}/{2}$.
Again, in the special case $\zomega_\B=\zomega_\F\,(\equiv\zomega_0)$, one finds the Kohn mode: $\Omega_+=\zomega_0$ and
$\Gamma_+=0$~\cite{kohn_1961_PR_123_cyclotron};
the other mode reduces to the damped oscillation mode: $\Omega_-=\sqrt{\zomega_0^2-\Delta/M_+}$ and $\Gamma_-=2/\tau$.

We observe the interesting features of the dipole modes from
Eqs.~\eqref{eq:eigenvalue_in_hydrodynamic_limit_in_dipole_mode}
and~\eqref{eq:eigenfrequency_and_damping_rate_in_collisionless_limit_in_dipole_mode}.
In the collisionless regime, there are two oscillating modes.
On the other hand, in the hydrodynamic regime, there are a single oscillating mode and two purely-damped modes.
In the transition from the collisionless regime to the hydrodynamic regime, one oscillating mode disappears and the two purely-damped modes emerge instead.
One is the fast relaxing mode: the damping rate of which is proportional to $1/\tau$.
This mode describes an out-of-phase motion between two components, immediately approaching static equilibrium.
The other relaxation mode is the slow relaxing mode: the damping rate of which is proportional to $\tau$.
This mode also involves the out-of-phase motion but exhibits very slow relaxation distinct from the fast relaxing mode.

\begin{figure}
\centering
\includegraphics[scale=.54]{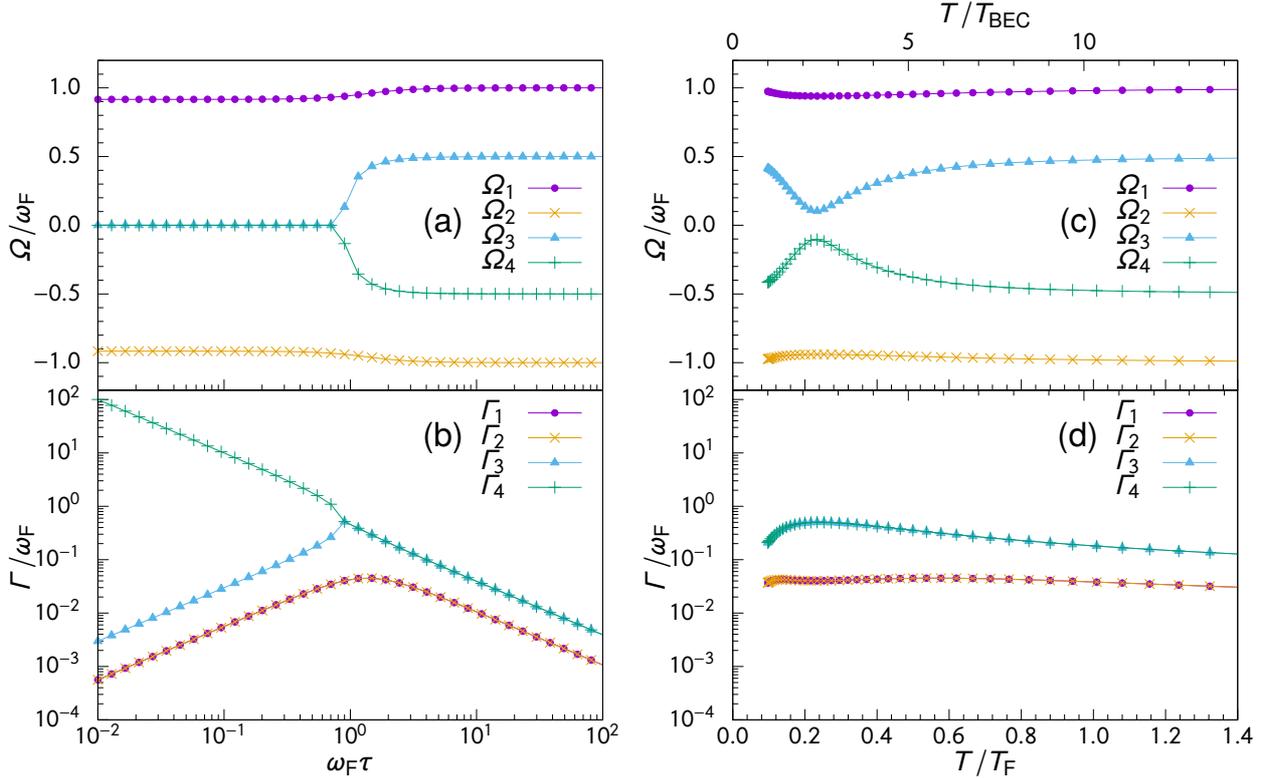}
\caption{Dipole mode $\omega_i=\Omega_i-\ui\Gamma_i\,(i=1,2,3,4)$ as a result of solving the coupled moment equation
in Eq.~\eqref{eq:dipole_moment_eq}
with the parameters $m_\B/m_\F=2.175$ (corresponding a mixture $^{87}\mathrm{Rb}$-$^{40}\mathrm{K}$), $N_\B/N_\F=0.125$, $\hbar\omega_\B/(\kB\TF)=0.005$, $\hbar\omega_\F/(\kB\TF)=0.01$, $g_\BB n^\ideal_\F/(\kB\TF)=g_\BF n^\ideal_\F/(\kB\TF)=0.1$.\;
\textbf{a}, \textbf{b} 
The left column is the result obtained by setting $\Delta=0$, plotting relaxation time dependence of frequency $\Omega_i$ and damping rate $\Gamma_i$.\;
\textbf{c}, \textbf{d} The right column is the result including $\Delta$, plotting temperature dependence of frequency $\Omega_i$ and damping rate $\Gamma_i$.
Temperature $\TF$ denotes Fermi temperature for the ideal Fermi gas given by $\kB\TF=\hbar\zomega_\F(6N_\F)^{1/3}$.
The data points plotted in \textbf{c} and \textbf{d} are for $T\geq T_\BEC$, where $T_\BEC$ denotes Bose--Einstein condensation temperature of the Bose--Fermi mixture gas.
Here, $T_\BEC$ includes contributions of the trap frequency and the mean-field terms.
A smaller number ratio $N_\B/N_\F$ makes the temperature ratio $T_\BEC/\TF$ smaller, and thus we can explore lower temperature region by setting $N_\B/N_\F=0.125$ (Color
figure online)
}
\label{fig:002490_val_01_00}%
\end{figure}

We also show the features of the dipole modes in the entire region including the intermediate region between the two limits.
Figure~\ref{fig:002490_val_01_00}
plots solutions of the coupled moment equation
in Eq.~\eqref{eq:dipole_moment_eq}.
Figure~\ref{fig:002490_val_01_00}a and~b
clearly exhibit the transition of collective modes between the two regimes at $\zomega_\F\tau\approx0.70$: disappearance of one oscillating mode and appearance of two purely-damped modes.
With respect to the temperature dependence, 
Fig.~\ref{fig:002490_val_01_00}c, d
does not show the transition for the set of parameters we used.
Two frequency $\Omega_3$ and $\Omega_4$ approach zero at $T/\TF\approx0.23$, where in fact the relaxation time shows a minimum.
In both the high temperature and low temperature regions, the two low-frequency modes with $\Omega_3$ and $\Omega_4$ correspond to the collisionless modes with a long relaxation time $\tau$
in Fig.~\ref{fig:002490_val_01_00}a and~b.
The tendency to become the collisionless regime in a high temperature region is due to the low particle density in trapped gases, leading to the small relaxation rate.
In the low temperature region, the relaxation rate becomes small again, because of suppression of the interspecies collisions due to Fermi degeneracy.
The system is close to the hydrodynamic regime at the temperature $T/\TF\approx0.23$, where $\Omega_3$ and $\abs*{\Omega_4}$ take the minimum values.
Since one does not reach the hydrodynamic regime, the damping rate
in Fig.~\ref{fig:002490_val_01_00}d
does not show the bifurcation to the two purely-damped modes.
By tuning the parameter set, one can realize the situation where the hydrodynamic regime appears in the intermediate temperature region, where the bifurcation to the two purely-damped modes emerges, which is reported in a separate
paper~\cite{asano_2020_PRA_101_dipole}.

\section{Conclusions}
\label{sec:conclusions}
We investigated dipole modes of a trapped Bose--Fermi mixture gas in the normal phase with $s$-wave scattering, composed of single-species bosons and single-species fermions.
Applying the moment method to the linearized Boltzmann equation, we studied the frequency and the damping rate in the hydrodynamic regime, collisionless regime, and intermediate regime between them as a function of the relaxation time.
Temperature dependence of these modes was also studied.
In the transition from the collisionless regime to the hydrodynamic regime, an oscillating mode disappears, and two purely-damped modes emerge.
One is the fast relaxing mode, and the other is the slow relaxing mode.
The latter is of interest, because its damping rate is so small that the two component are gradually mixed through the out-of-phase motion even in the hydrodynamic regime, where the relaxation time is very small.
Considering the temperature dependence of the relaxation time, the transition to the hydrodynamic regime does not achieve for the set of parameters chosen in the present paper.
We found the tendency to become the collisionless regime not only in the high temperature region but also in the extremely low temperature region because of Fermi degeneracy.
In a separate paper, we address the transition of the eigenmodes, and discuss physics of the slow relaxation mode in
detail~\cite{asano_2020_PRA_101_dipole}.
Although we focused on a mixture gas in the normal phase composed of single-species bosons and single-species fermions, it will be interesting future issues to analyze the dipole modes in the superfluid phase, and in a Bose--Fermi mixture gas with two-species fermions.

\section*{Acknowledgements}
S.~W.~was supported by JSPS KAKENHI Grant No.~JP18K03499, and T.~N.~was supported by JSPS KAKENHI Grant No.~JP16K05504.

\bibliographystyle{spphys}
\bibliography{qfs2019pro}
\end{document}